\documentclass[10pt,pra,twocolumn,amssymb,nofootinbib,showpacs]{revtex4}
\usepackage{bm}
\usepackage{graphicx}
\usepackage{amsmath}
\usepackage{amssymb}
\usepackage{bbm}

\begin{document}
\title{\bf Geometric phase for open quantum systems and stochastic
unravellings}
\author{Angelo Bassi}
\email{bassi@ictp.triste.it} \affiliation{Mathematisches Institut,
Universit\"at M\"unchen, Theresienstr. 39, 80333 M\"unchen, Germany}
\affiliation{The Abdus Salam International Centre for Theoretical
Physics, Strada Costiera 11, 34014 Trieste, Italy}
\author{Emiliano Ippoliti}
\email{ippoliti@ts.infn.it} \affiliation{Department of Theoretical
Physics, University of Trieste, Strada Costiera 11, 34014 Trieste,
Italy} \affiliation{Istituto Nazionale di Fisica Nucleare, sezione
di Trieste, Via Valerio 2, 34127 Trieste, Italy}
\begin{abstract}
We analyze the geometric phase for an open quantum system when
computed by resorting to a stochastic unravelling of the reduced
density matrix (quantum jump approach or stochastic Schr\"odinger
equations). We show that the resulting phase strongly depends on
the type of unravelling used for the calculations: as such, this
phase is {\it not} a geometric object since it depends on
non-physical parameters which are not related to the path followed
by the density matrix during the evolution of the system.
\end{abstract}
 \maketitle

\section{introduction}

The geometric phase is a property of a physical system which
depends only on the path the system follows during its evolution,
not on the details of the dynamics. Since the work of M. Berry
\cite{ber}, geometric phases have acquired a primary role in our
understanding of many physical phenomena \cite{lib}, and have been
subject to several experimental verifications \cite{exp}. The
original idea, framed within the context of adiabatic and cyclic
evolutions of isolated systems, has been generalized in various
directions \cite{gen1, gen2, gen3, gen3.1, new}; of particular
importance are all those efforts \cite{gen2, gen3, gen3.1, new,
gen4} aiming at defining a geometric phase for open quantum
systems, which are not mathematically described by a pure state
$|\psi_{t}\rangle$ but in terms of a reduced density matrix
$\rho_{t}$ undergoing a non-unitary evolution. Such proposals,
besides been interesting on their own, could be important for the
possible applications, e.g., to quantum computation \cite{qc1}.

The mathematical framework within which the geometric phase for an
open quantum system is defined is the following:
\\ \\ \noindent
1. One assumes that the effect of the environment on the quantum
system is such that, under suitable approximations, the system can
be {\it effectively} treated as an isolated system undergoing a
non-unitary type of {\it linear}\footnote{As discussed in ref.
\cite{gisfl2} the evolution equation for the statistical operator
must be linear, otherwise it can give origin to superluminal
effects.} evolution:
\begin{equation} \label{eq:map}
\Sigma_{t}: \rho_{0} \; \rightarrow \; \Sigma_{t} [ \rho_{0} ]
\equiv \rho_{t},
\end{equation}
which takes into account both the internal dynamics of the system
and its interaction with the environment. Under reasonable
assumptions \cite{Lindblad}, which anyhow have been questioned in
the literature (see e.g. \cite{bas}), the map $\Sigma_{t}$ can be
taken of the quantum-dynamical-semigroup type, generated by the
following class of equations:
\begin{equation} \label{efr}
\frac{d\rho}{dt} = -\frac{i}{\hbar}\, \left[ H, \rho \right] -
\frac{\lambda^2}{2} \sum_{n = 1}^{N} \left\{ L^{\dagger}_{n} L_{n}
\rho + \rho L^{\dagger}_{n} L_{n} -2 L_{n} \rho L^{\dagger}_{n}
\right\};
\end{equation}
the self--adjoint operator $H$ is usually identified with the
standard Hamiltonian of the system, while the operators $L_{n}$,
together with the positive constant $\lambda$, summarize the
effect of the environment on the system.

A general consequence of this type of approach is that a pure
state $|\psi_{t}\rangle$ is usually mapped into a statistical
mixture $\rho_{t}$, so the problem arises of how to identify a
geometric phase for the evolution of a density matrix $\rho_{t}$.
\\ \\ \noindent
2. A common strategy which is used in the literature for
associating a geometric phase to the evolution of $\rho_{t}$ is to
formally map the density matrix into a statistical mixture of pure
states $|\psi_{t}^{n}\rangle$, each of which is weighted with a
probability $p_{n}(t)$:
\begin{equation} \label{eq:rho-ensemble}
\rho_{t} \rightarrow \{ (|\psi_{t}^{n}\rangle, p_{n}(t) )\}:
\qquad \rho_{t} = \sum_{n} p_{n}(t) |\psi_{t}^{n}\rangle\langle
\psi_{t}^{n}|;
\end{equation}
one can then use the standard definition of geometric phase for a
pure state:
\begin{eqnarray} \label{gf}
\gamma^{\makebox{\tiny geo}}_{t} & = & \gamma^{\makebox{\tiny
tot}}_{t} \; - \;
\gamma^{\makebox{\tiny dyn}}_{t} \nonumber \\
& = & \makebox{Arg}\;\langle\psi_{0}|\psi_{t}\rangle -
\makebox{Im} \int_{0}^{t} \langle\psi_{t}| d |\psi_{t}\rangle,
\end{eqnarray}
to associate a geometric phase also to $\rho_{t}$; this strategy
as given fruitful results in the case of mixed states undergoing a
unitary evolution \cite{gen3}, while the case of non-unitary
evolutions, in particular those associated to open quantum
systems, is still under debate. In this second case, a tentative
definition of geometric phase has been given via {\it state
purification} \cite{gen3.1} and the {\it quantum-jump approach}
\cite{gen4}.

A well-known characteristic property of
relation~\eqref{eq:rho-ensemble} is that the association between a
density matrix $\rho_{t}$ and an ensemble $\{
(|\psi_{t}^{n}\rangle, p_{n}(t) )\}$ is not one-to-one, but
one-to-many \cite{NC}: in general, there are different ensembles,
containing different vectors and different probabilities, which
give rise to the same density matrix; moreover, such vectors and
the corresponding probabilities evolve in completely different
ways. In the state purification approach, this property of density
matrices is related to the fact that there are different Kraus
representations of the dynamical evolution of $\rho_{t}$
\cite{Davis}. In the quantum-jump approach, the same property
reflects the fact that there are different stochastic unravellings
which end up to reproduce the same statistical operator
\cite{Carmichael}.

In this paper, we analyze the consequences of such a feature of
density matrices for the definition of geometric phases, within
the context of the stochastic unravelling formalism, which the
quantum-jump approach belongs to. We will show that the approach
followed in \cite{gen4} to identify a geometric phase strongly
depends on the type of unravelling of $\rho_{t}$: as a
consequence, such a phase is not a geometric object, because it
depends also on non-physical parameters which are not related to
the path followed by the density matrix during its evolution.

\section{stochastic unravellings}

A stochastic unravelling of a linear evolution $\Sigma_{t}$ for a
density matrix $\rho_{t}$ is defined as follows. Let us fix a
probability space $(\Omega, {\mathcal F}, {\mathbb  P})$ and let
us consider a stochastic evolution for statevectors:
\begin{equation} \label{eq:soch-ev}
T_{t}(\omega): |\psi_{0}\rangle \; \rightarrow
|\psi_{t}(\omega)\rangle,
\end{equation}
which, for each different sample element $\omega \in \Omega$,
associates a different statevector $|\psi_{t}(\omega)\rangle$ to
the same initial state $|\psi_{0}\rangle$. One can then define the
density matrix:
\begin{equation} \label{eq:stoch-def}
\tilde\rho_{t} \; \equiv \; \sum_{n} p_{n}\, {\mathbb E}_{\mathbb P}
[ |\psi^{n}_{t}(\omega)\rangle\langle\psi^{n}_{t}(\omega)| ],
\end{equation}
where the symbol ${\mathbb E}_{\mathbb P}$ denotes the average
value with respect to the probability measure ${\mathbb P}$; here
above we have assumed that the initial state of the system is
represented by a statistical mixture $\{ (|\psi_{0}^{n}\rangle,
p_{n}) \}$, so an extra sum over the possible initial states
appears in the definition of $\tilde\rho_{t}$. The above relation
defines a map:
\begin{equation} \label{eq:tilde-map}
\tilde\Sigma_{t} : \rho_{0} \; \rightarrow \; \tilde\Sigma_{t} [
\rho_{0} ] \equiv \tilde\rho_{t},
\end{equation}
where $\tilde\rho_{t}$ is given by Eq.~\eqref{eq:stoch-def}. Now,
if the map $\tilde\Sigma_{t}$ defined by Eqs.~\eqref{eq:tilde-map}
and~\eqref{eq:stoch-def} coincides with the linear map
$\Sigma_{t}$, we say that $T_{t}(\omega)$ is a {\it stochastic
unravelling} of $\Sigma_{t}$.

Among the other things, the above definition of stochastic
unravelling implies that, when computing observable quantities of
a system which evolves according to the map $\Sigma_{t}$, e.g. the
expectation value of a self-adjoint operator: $\langle O
\rangle_{t} = \makebox{Tr}[ O \rho_{t}]$, one can start with a
stochastic unravelling~\eqref{eq:soch-ev} of $\Sigma_{t}$, then he
computes the quantum expectation
$\langle\psi_{t}(\omega)|O|\psi_{t}(\omega)\rangle$ and finally he
averages over the noise; this sequence of operations is legitimate
since, by the definition of stochastic unravelling, one trivially
has:
\begin{equation} \label{eq:aver-val}
{\mathbb E}_{\mathbb P} [ \langle \psi_{t}|O|\psi_{t} \rangle ] \;
= \; \makebox{Tr}[ O \Sigma_{t}[|\psi_{0}\rangle\langle \psi_{0}|]
];
\end{equation}
of course, if the initial state is a mixed state, an extra sum
over the possible initial states, weighted with the corresponding
probability distribution, has to be added at the left hand side
of~\eqref{eq:aver-val}.

In the literature, two types of stochastic unravellings have been
proposed, one discrete and one continuous. The first one is the
{\it quantum jump approach} \cite{Carmichael} which has been used
in refs. \cite{gen4} to associate a geometric phase to the
evolution of an open quantum system; the second one is given in
terms of {\it stochastic Schr\"odinger equations}
\cite{barchielli}. The two approaches are similar and, in the
following, we will resort to the second one, because it is more
elegant from the mathematical point of view and easy to handle. We
now briefly review it.

The idea is simple: the stochastic evolution $T_{t}(\omega)$ is
assumed to be generated by a stochastic Schr\"odinger equation,
whose typical structure is the following
\cite{barchielli,arn,gard}:
\begin{eqnarray} \label{nlin}
d |\psi_{t}\rangle & = & \left[ - \frac{i}{\hbar} H dt +
\lambda\sum_{n =
1}^{N} (L_{n} - r_{n,t})\, dW^{n}_{t} \right. \\
& & \left. - \frac{\lambda^2}{2} \sum_{n = 1}^{N}
(L^{\dagger}_{n}L_{n} -2L_{n}r_{n,t} + r_{n,t}^{2})\, dt \right]
|\psi_{t}\rangle, \nonumber
\end{eqnarray}
where:
\begin{equation}
r_{n,t} \; = \; \frac{1}{2}\, \langle \psi_{t}| [ L^{\dagger}_{n}
+ L_{n} ] |\psi_{t}\rangle,
\end{equation}
with $H$ and $L_{n}$ defined as in~\eqref{efr}; $W^{n}_{t}$ ($n =
1, \ldots, N$) are $N$ independent standard Wiener processes with
respect to the measure ${\mathbb P}$, which make Eq.~\eqref{nlin}
a stochastic differential equation.

Such kind of equations have been used in several contexts: within
the {\it theory of quantum measurement}, to describe the effects
of a repeated measurement on the evolution of a quantum system
\cite{qm}; within {\it collapse models}, to provide a solution to
the measurement problem \cite{cm}; within the theory of {\it open
quantum system}, as a mathematical tool to efficiently simulate
the evolution of an open system \cite{dec}.

One of the fundamental properties \cite{barchielli} of
Eq.~\eqref{nlin} is that the density matrix  $\rho_{t} \equiv
{\mathbb E}_{\mathbb P} [ |\psi_{t} \rangle\langle \psi_{t}| ]$
solves Eq.~\eqref{efr}, i.e. Eq.~\eqref{nlin} represents a
stochastic unravelling of the Lindblad-type equation~\eqref{efr}.

Note that Eq.~\eqref{nlin} is non-linear, but it preserves the
norm of the statevector. There is a well-known way
\cite{barchielli} to linearize the equation, at the price of
relinquishing the normalization condition; consider the following
stochastic differential equation:
\begin{equation} \label{eq:lin}
d |\phi_{t}\rangle \! = \! \left[ - \frac{i}{\hbar} H dt +
\lambda\sum_{n = 1}^{N} L_{n}\, d\xi^{n}_{t} - \frac{\lambda^2}{2}
\sum_{n = 1}^{N} L^{\dagger}_{n}L_{n}\, dt \right] \!
|\phi_{t}\rangle;
\end{equation}
the stochastic processes $\xi^{n}_{t}$ are standard Wiener
processes with respect to a new probability measure ${\mathbb Q}$,
whose relation to ${\mathbb P}$ will soon be established.

The connection between the linear Eq.~\eqref{eq:lin} and the
nonlinear Eq.~\eqref{nlin} is the following; given the solution
$|\phi_{t}\rangle$ of Eq.~\eqref{eq:lin} for a initial condition
$|\phi_{0}\rangle$, if one performs the following two operations:
\begin{enumerate}

\item Normalize the solution: $|\phi_{t}\rangle \rightarrow
|\psi_{t}\rangle = |\phi_{t}\rangle/\| |\phi_{t}\rangle \|$,

\item Make the substitution:
\begin{equation} \label{eq:nois-sub}
\xi^{n}_{t} \; \longrightarrow \; W^{n}_{t} \; = \; \xi^{n}_{t} -
2 \lambda \int_{0}^{t} r_{n,t}\; ds,
\end{equation}
\end{enumerate}
then the wavefunction $|\psi_{t}\rangle$ so defined is the
solution of Eq.~\eqref{eq:spin-eq} for the same initial condition
$|\psi_{0}\rangle = |\phi_{0}\rangle$. Moreover, one can further
show that the two probability measures ${\mathbb P}$ and ${\mathbb
Q}$ are related as follows \cite{barchielli}:
\begin{equation} \label{eq:rel-lin-nlin}
{\mathbb E}_{\mathbb P} [X_{t}] \; \equiv \; {\mathbb E}_{\mathbb
Q} [ \langle \phi_{t} | \phi_{t} \rangle X_{t} ],
\end{equation}
where $X_{t}$ is a stochastic process.

\subsection{Equivalent stochastic unravellings}

Two stochastic unravellings $T^{(1)}_{t}$ and $T^{(2)}_{t}$ are said
to be {\it equivalent} if they unravel the same evolution
$\Sigma_{t}$. A very remarkable property is that there are {\it
infinite} different but equivalent stochastic unravellings for
practically all physically interesting $\Sigma_{t}$; within the
quantum jump approach, this issue is addressed, e.g., in
\cite{Carmichael}; within the stochastic Schr\"odinger formalism,
such a feature is less known, still very easy to show. As a matter
of fact, suppose in Eq.~\eqref{nlin} we change the Lindblad
operators $L_{n}$ as follows:
\begin{equation} \label{eq:phase-change}
L_{n} \; \longrightarrow c_{n}\, L_{n},
\end{equation}
where $c_{n} = e^{i \varphi_{n}}$ are arbitrary phase factors;
clearly, Eq.~\eqref{efr} does not change, while Eq.~\eqref{nlin}
does change, since terms appear which are not proportional to
$L^{\dagger}_{n} L_{n}$. Such a change is not as trivial as it may
seem: as we shall see, stochastic equations with different values
of $\varphi_{n}$ entail completely different evolutions for the
statevector. Of course, there are other possible unravellings of
Eq.~\eqref{efr}, besides those which can be obtained by a phase
shift in the Lindblad operators, but for simplicity we consider
here only these, since they are sufficient for the subsequent
analysis.

\section{geometric phase and stochastic unravellings: an example}

As already remarked, in refs. \cite{gen4} the
stochastic-unravelling approach has been used to associate a
geometric phase to an open system; in this section we discuss how
the existence of different equivalent stochastic unravellings
affects the computation of the geometric phase. As an example, we
now calculate the geometric phase associated to the evolution of a
spin particle in a constant magnetic field directed along the
$z$--axis of a chosen reference frame, while the spin is subject
to dephasing. The quantum Hamiltonian is $H = -\mu B \sigma_{z}$,
and the effect of the environment is described by one Lindblad
operator $L = \sigma_{z}$; the corresponding Lindblad equation is
($\hbar = 1$):
\begin{equation} \label{eq:lind}
\frac{d}{dt} \rho_{t} = i \mu B \left[ \sigma_{z}, \rho_{t}
\right] - \frac{\lambda^2}{2} \left[ \sigma_{z}, \left[
\sigma_{z}, \rho_{t} \right] \right].
\end{equation}
The initial spin state is taken equal to:
\begin{equation} \label{ic}
|\psi_{0}\rangle \; = \; \cos\frac{\theta}{2} |+\rangle \, + \,
\sin\frac{\theta}{2} |-\rangle,
\end{equation}
where $|+\rangle$ and $|-\rangle$ are the two eigenstates of
$\sigma_{z}$. Eq.~\eqref{nlin} becomes:
\begin{eqnarray} \label{eq:spin-eq}
d |\psi_{t}\rangle & = & \Big[ i \mu B \sigma_{z} dt + \lambda ( c
\sigma_{z} - \cos\varphi\, \langle \sigma_{z} \rangle_{t}) dW_{t} \\
& - & \left. \frac{\lambda^2}{2} ( \sigma_{z}^2 - 2c \cos\varphi\,
\langle \sigma_{z} \rangle_{t}\, \sigma_{z} + \cos^2\!\varphi\,
\langle \sigma_{z} \rangle_{t}^2 ) \right] |\psi_{t}\rangle,
\nonumber
\end{eqnarray}
with $\langle \sigma_{z} \rangle_{t} = \langle \psi_{t}|
\sigma_{z} |\psi_{t} \rangle$. In the above equation, we have
included also the arbitrary phase factor $c = e^{i\varphi}$ which,
as already discussed, does not appear in Eq.~\eqref{eq:lind} for
the density matrix $\rho_{t}$. We now compute the total and
dynamical phases associated to the ensemble of vectors
$\{|\psi_{t}\rangle \equiv |\psi_{t}(\omega)\rangle, \omega \in
\Omega \}$ generated by Eq.~\eqref{eq:spin-eq}.

\subsection{Total phase}

To derive the correct formula for the total phase, we resort to an
interferometric scheme like the one depicted in Fig.~1, i.e. a
Mach-Zehnder interferometer with a variable phase shifter $\chi$
in one of the two arms and the magnetic field in the other.
\begin{figure}[t!]
\setlength{\unitlength}{0.7pt}
\begin{picture}(300,140)(0,40)
\put(0,40){\line(1,0){300}} \put(0,200){\line(1,0){300}}
\put(0,40){\line(0,1){160}} \put(300,40){\line(0,10){160}}
\thicklines
\put(20,160){\line(1,0){113}} \put(157,160){\line(1,0){73}}
\put(20,160){\line(1,0){40}} \put(60,160){\line(0,-1){30}}
\put(58,80){\line(1,0){210}} \put(230,50){\line(0,1){110}}
\put(58,110){\line(0,-1){30}}
\put(50,170){\line(1,-1){20}} {\thicklines
\put(48,90){\line(1,-1){20}}
\put(46,88){\line(1,-1){20}} \put(221,170){\line(1,-1){20}}
\put(223,172){\line(1,-1){20}}} \put(220,90){\line(1,-1){20}}
\put(30,157){$\blacktriangleright$}
\put(250,77.5){$\blacktriangleright$}
\put(225.5,54){$\blacktriangledown$}
\put(145,160){\circle{25}} \put(50,110){\framebox(20,20){\large B}}
\put(140,158){\large $\chi$} \put(239,90){\large
$|\makebox{h}\rangle$} \put(200,55){\large $|\makebox{v}\rangle$}
\end{picture}\caption{Interferometric scheme for measuring the total phase.
$|{\mathrm h}\rangle$ corresponds to the beam travelling in the
horizontal direction, while $|{\mathrm v}\rangle$ corresponds to
the beam travelling in the vertical direction. $\chi$ is a
variable phase shifter and $B$ a magnetic field.}
\end{figure}
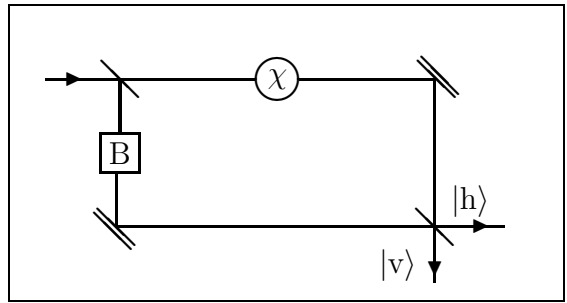
We call $|\makebox{h}\rangle$ and $|\makebox{v}\rangle$ the
spatial part of the wavefunction when the beam travels along the
horizontal and vertical direction, respectively; in the subsequent
analysis, we will neglect the contribution to the phase given by
the free part of the evolution, since we assume that the length of
the two arms of the interferometer is the same. The initial state
of the beam then is:
\begin{equation}
|\Psi_{0}\rangle = |\psi_{0}\rangle \otimes |\makebox{h}\rangle \;
= \; \left[ \cos\frac{\theta}{2} |+\rangle \, + \,
\sin\frac{\theta}{2} |-\rangle \right] \otimes
|\makebox{h}\rangle.
\end{equation}
To find the evolution from the initial time $t = 0$ to the final
time $t = t_{\makebox{\tiny F}}$ when the beam comes out through
the interferometer, we first consider the linear version of
Eq.~\eqref{eq:spin-eq}, which is\footnote{In the following when we
write $|\Psi_{t}\rangle$ and $|\Phi_{t}\rangle$, we mean that
Eqs.~\eqref{eq:spin-eq} and~\eqref{eq:lin-eq} include not only the
spin degree of freedom, but also the spatial one; when on the
other hand we write $|\psi_{t}\rangle$ and $|\phi_{t}\rangle$, we
mean that we are taking into account only the spin degree of
freedom.}:
\begin{equation} \label{eq:lin-eq}
d |\Phi_{t}\rangle \; = \; \left[ i\mu B \sigma_{z} dt \, + \,
\lambda c \sigma_{z} d\xi_{t} \, - \, \frac{\lambda^2}{2}
\sigma_{z}^2 dt \right] |\Phi_{t}\rangle.
\end{equation}
By taking into account the effects of the two mirrors and partial
beam splitters, one finds for the beam at time $t_{\makebox{\tiny
F}}$:
\begin{eqnarray}
|\Phi_{t_{\makebox{\tiny F}}}\rangle & = & \frac{i}{2} \left[ \left(
e^{i \chi} + e^{i \mu B T} \right) e^{\lambda c
\xi_{t_{\makebox{\tiny F}}} -
\frac{\lambda^2}{2} (1+c^2)t_{\makebox{\tiny F}}}\cos\frac{\theta}{2}\,|+\rangle \right. \nonumber \\
& + & \!\!\left. \left( e^{i \chi} + e^{-i \mu B T} \right)
e^{-\lambda c \xi_{t_{\makebox{\tiny F}}} - \frac{\lambda^2}{2}
(1+c^2)t_{\makebox{\tiny F}}}\sin\frac{\theta}{2}\,|-\rangle \right]
|\makebox{h}\rangle \nonumber
\\
& + &\!\! \frac{1}{2} \left[  \left( e^{i \chi} - e^{i \mu B T}
\right) e^{\lambda c \xi_{t_{\makebox{\tiny F}}} -
\frac{\lambda^2}{2} (1+c^2)t_{\makebox{\tiny F}}}\cos\frac{\theta}{2}\,|+\rangle \right. \nonumber \\
& + & \!\!\left. \left( e^{i \chi} - e^{-i \mu B T} \right)
e^{-\lambda c \xi_{t_{\makebox{\tiny F}}} - \frac{\lambda^2}{2}
(1+c^2)t_{\makebox{\tiny F}}}\sin\frac{\theta}{2}\,|-\rangle \right]
|\makebox{v}\rangle \nonumber
\\
\end{eqnarray}

One can now compute the output intensity along
$|\makebox{h}\rangle$:
\begin{eqnarray}
I_{t_{\makebox{\tiny F}}}({\omega}) & = & \langle
\Psi_{t_{\makebox{\tiny F}}}| \left[
|\makebox{h}\rangle\langle\makebox{h}| \otimes {\mathbb
I}_{\makebox{\tiny SPIN}} \right] | \Psi_{t_{\makebox{\tiny F}}}
\rangle \label{eq:int} \\
& = & \frac{1}{2} + \frac{1}{2}\,| f_{t_{\makebox{\tiny
F}}}({\omega}) |\, \cos \left( \chi + \makebox{Arg}\left(
f_{t_{\makebox{\tiny F}}}({\omega}) \right) \right), \nonumber
\end{eqnarray}
where $|\Psi_{t_{\makebox{\tiny F}}}\rangle =
|\Phi_{t_{\makebox{\tiny F}}}\rangle / \| |\Phi_{t_{\makebox{\tiny
F}}}\rangle \|$ and:
\begin{eqnarray} \label{eq:fom}
f_{t_{\makebox{\tiny F}}}({\omega}) & = &  \frac{e^{-2\lambda^2
t_{\makebox{\tiny F}}\cos^2\!\varphi}}{\langle
\Phi_{t_{\makebox{\tiny F}}}| \Phi_{t_{\makebox{\tiny F}}}
\rangle} \left[ e^{2 \lambda \xi_{t_{\makebox{\tiny F}}} \cos
\varphi - i \mu B T} \cos^2\!
\frac{\theta}{2}\right.\nonumber \\
& + & \left. e^{-2 \lambda \xi_{t_{\makebox{\tiny F}}} \cos
\varphi + i \mu B T} \sin^2\!\frac{\theta}{2}\right].
\end{eqnarray}
One can then identify the total phase, for each realization of the
noise, as:
\begin{equation} \label{eq:tot-ph}
\gamma^{\makebox{\tiny tot}}_{t}({\omega}) \equiv \makebox{Arg}[
f_{t}({\omega})],
\end{equation}
which depends not only on $\omega$, but also on $\varphi$, i.e. on
the type of unravelling of the master equation~\eqref{eq:lind}.
Anyway, $\gamma^{\makebox{\tiny tot}}_{t}({\omega})$ as such is
{\it not} a physical quantity because it cannot be observed. The
final outcome---i.e. the interference pattern---consists of many
repetitions of the experiment, accordingly the observable physical
quantity is the {\it average} intensity $I_{t} \equiv {\mathbb
E}_{\mathbb P}[ I_{t}({\omega}) ]$, which can be easily computed
by taking into account relation~\eqref{eq:rel-lin-nlin}:
\begin{equation}
I_{T} \; = \; \frac{1}{2} \, + \, \frac{1}{2}\, \nu_{T}\, \cos
(\chi - \gamma^{\makebox{\tiny tot}}_{T}),
\end{equation}
where:
\begin{equation}
\nu_{T} \; \equiv \; | {\mathbb E}_{\mathbb P}[
f_{t_{\makebox{\tiny F}}}({\omega}) ]| \; = \; \left| e^{i \mu B
T} \cos^2\!\frac{\theta}{2}+ e^{-i \mu B
T}\sin^2\!\frac{\theta}{2} \right|
\end{equation}
is the output visibility, while:
\begin{eqnarray} \label{eq:tot-phase}
\gamma^{\makebox{\tiny tot}}_{T} & = & \makebox{Arg} {\mathbb
E}_{\mathbb P}[ f_{t_{\makebox{\tiny
F}}}({\omega}) ] \nonumber \\
& = & \makebox{Arg}\, \left[ e^{i \mu B T}
\cos^2\!\frac{\theta}{2}+
e^{-i \mu B T}\sin^2\!\frac{\theta}{2}\right] \nonumber \\
& = & \makebox{Arg}\; {\mathbb E}_{\mathbb Q} [ \langle \phi_{0} |
\phi_{t} \rangle ],
\end{eqnarray}
is the total phase difference; in particular, if $B$ acts on the
spin for a time $T = \pi/\mu B$, we have the standard result:
$\gamma^{\makebox{\tiny tot}}_{T} = \pi$.

Note that the total phase $\gamma^{\makebox{\tiny tot}}_{T}$ does
not depend on $\varphi$, i.e. it does not depend on the specific
choice of the unravelling used to make the calculations. Such a
result is a consequence of property~\eqref{eq:aver-val} of
stochastic unravellings, i.e. of the fact that Eq.~\eqref{eq:int}
for the output intensity, when also the average over the noise is
taken into account, can be expressed as a function of the density
matrix ($I_{t} \equiv {\mathbb E}_{\mathbb P}[ I_{t}({\omega}) ] =
\makebox{Tr} [ (|\makebox{h}\rangle\langle\makebox{h}| \otimes
{\mathbb I}_{\makebox{\tiny SPIN}}) \rho_{t} ]$), so that any
dependence on the unravelling disappears. Such a result is then
not a peculiar byproduct of the specific model taken into account,
but a {\it necessary} mathematical consequence of the formalism.
This fact can be seen in a different way: the total phase
difference between the two arms of the interferometer is a
physical quantity which can be experimentally measured; like all
physical quantities, it must be deducible from the master
equation~\eqref{eq:lind}, so it does not have to depend on
$\varphi$.

As a final observation, we note that both the average visibility
and the average total phase do not depend on $\lambda$. This
specific fact is a consequence of our simple model of open quantum
system, according to which the noise is perfectly correlated among
the two arms of the interferometer. Of course such an assumption
is not realistic, and it has been made only to simplify the
calculations, since it does not affect the conclusion of our work.

\subsection{Dynamical phase}

We now compute the dynamical phase $\gamma^{\makebox{\tiny
dyn}}_{t}$ induced by the precession of the spin-system when
interacting with the magnetic field; by using It\^o calculus
\cite{arn}, one finds from Eq.~\eqref{eq:spin-eq}:
\begin{eqnarray}
\langle\psi(t)| d |\psi(t)\rangle & = & i\mu B \langle \sigma_{z}
\rangle_{t}\, dt + i \lambda \sin\varphi \, \langle \sigma_{z} \rangle_{t}
dW_{t} \\
& - & \frac{\lambda^2}{2}\left[ 1 - (2 c \cos\varphi -
\cos^2\!\varphi)\langle \sigma_{z} \rangle_{t}^{2}\right] dt,
\nonumber
\end{eqnarray}
where only the imaginary part has to be taken into account. For
each realization of the noise, the dynamical phase is:
\begin{eqnarray} \label{eq:din-phase}
\gamma^{\makebox{\tiny dyn}}_{t}({\omega}) & = & \mu B
\int_{0}^{t} \langle \sigma_{z} \rangle_{s}\, ds + \lambda
\sin\varphi\! \int_{0}^{t}
\langle \sigma_{z} \rangle_{s}\, dW_{s} \nonumber \\
& & + \lambda^2 \sin\varphi\cos\varphi \int_{0}^{t} \langle
\sigma_{z} \rangle_{s}^2 \, ds,
\end{eqnarray}
which, like the total phase $\gamma^{\makebox{\tiny
tot}}_{t}({\omega})$, depends not only on $\omega$, but also on
the unravelling of the master equation.

We now compute the stochastic average $\gamma^{\makebox{\tiny
dyn}}_{t}$ of $\gamma^{\makebox{\tiny dyn}}_{t}(\omega)$, for
which we need to know the statistical properties of both $\langle
\sigma_{z} \rangle_{t}$ and $\langle \sigma_{z} \rangle_{t}^2$:
these can be easily computed by writing the corresponding
stochastic differential equations, both of which can be quite
easily derived from Eq.~\eqref{eq:spin-eq}. The equation for
$\langle \sigma_{z} \rangle_{t}$ is:
\begin{equation} \label{eq:sigma-z}
d\langle \sigma_{z} \rangle_{t} \; = \; 2\lambda
\cos\varphi\left[1 - \langle \sigma_{z} \rangle_{t}^{2}\right]
dW_{t},
\end{equation}
which tells us that since the Brownian increment $dW_{t}$ has zero
mean, the average value of $\langle \sigma_{z} \rangle_{t}$ does
not change in time: ${\mathbb E}_{\tiny\mathbb P} [\langle
\sigma_{z} \rangle_{t}] = {\mathbb E}_{\tiny\mathbb P} [\langle
\sigma_{z} \rangle_{0}] = \cos\theta$.

The stochastic differential equation for $\langle \sigma_{z}
\rangle_{t}^2$ instead is:
\begin{eqnarray} \label{eq:sigma-z-square}
d\langle \sigma_{z} \rangle_{t}^2 & = & 4 \lambda \cos\varphi \langle
\sigma_{z} \rangle_{t} \left[ 1 - \langle \sigma_{z} \rangle_{t}^2
\right] dW_{t}\nonumber \\
& + & 4 \lambda^2 \cos^2\!\varphi \left[ 1 - \langle \sigma_{z}
\rangle_{t}^2 \right ]^2 dt;
\end{eqnarray}
the first term on the r.h.s. does not contribute to the stochastic
average, so one has:
\begin{equation} \label{eq:sigma-z-square-average}
\frac{d}{dt}\, {\mathbb E}_{\mathbb P} [\langle \sigma_{z}
\rangle_{t}^2] \; = \; 4 \lambda^2 \cos^2\!\varphi\, {\mathbb
E}_{\mathbb P} \left[ 1 - \langle \sigma_{z} \rangle_{t}^2 \right
]^2 \; \geq \; 0,
\end{equation}
which implies that ${\mathbb E}_{\mathbb P} [\langle \sigma_{z}
\rangle_{t}^2]$ constantly increases in time, and in general
(unless $\varphi = \pi/2 + k \pi, k \in {\bf Z}$) it stops
increasing only when ${\mathbb E}_{\mathbb P} \left[ 1 - \langle
\sigma_{z} \rangle_{t}^2 \right ]^2 = 0$, i.e. when $\langle
\sigma_{z} \rangle_{t}^2 = 1$, with the possible exception of a
set of points $\omega \in \Omega$ of measure 0.

Concluding, the average dynamical phase $\gamma^{\makebox{\tiny
dyn}}_{t} \equiv {\mathbb E}_{\mathbb P}[ \gamma^{\makebox{\tiny
dyn}}_{t}({\omega})]$, after a time $T$, is equal to:
\begin{equation} \label{eq:geom-phase}
\gamma^{\makebox{\tiny dyn}}_{T}\; = \; \mu B T \cos\theta +
\lambda^2 \sin\varphi\cos\varphi \int_{0}^{T} {\mathbb E}_{\mathbb
P} [ \langle \sigma_{z} \rangle_{t}^2 ]\, dt,
\end{equation}
which, contrary to what happens to the average total phase, still
depends on $\varphi$, i.e. on the specific stochastic unravelling
of the master equation.

One could argue that, when computing the average dynamical phase,
we should not average over the phase, but over the phase factor,
i.e. we should compute ${\mathbb E}_{\mathbb P}[ \exp (
i\gamma^{\makebox{\tiny dyn}}_{t}({\omega}))]$ in place of
${\mathbb E}_{\mathbb P}[ \gamma^{\makebox{\tiny
dyn}}_{t}({\omega})]$, and then extract the argument; in this way
we would take into account the fact that a phase is defined
modulus $2 \pi$. The stochastic differential of $\exp (
i\gamma^{\makebox{\tiny tot}}_{t}({\omega}))$ is:
\begin{eqnarray}
d\, e^{i\gamma^{\makebox{\tiny tot}}_{t}({\omega})} & = & \Big[ i
\mu B \langle \sigma_{z} \rangle_{t} + i \lambda^2 \sin\varphi
\cos\varphi \langle \sigma_{z} \rangle_{t}^2 \nonumber \\
& & \left. - \frac{\lambda^2}{2} \sin^2 \varphi \langle \sigma_{z}
\rangle_{t}^2 \right] e^{i\gamma^{\makebox{\tiny
tot}}_{t}({\omega})}
dt \nonumber \\
& & + \left[ i \lambda \sin\varphi \langle \sigma_{z} \rangle_{t}
\right] e^{i\gamma^{\makebox{\tiny tot}}_{t}({\omega})} dW_{t}.
\end{eqnarray}
Its average value cannot be explicitly computed, due to the
dependence of both $\langle \sigma_{z} \rangle_{t}$ and $\langle
\sigma_{z} \rangle_{t}^2$ on the noise; anyway, when taking the
average, the dependence on $\varphi$ in general does not
disappear. E.g., if we take the trivial case in which the initial
state is $|\psi_{0}\rangle = |+\rangle$, so that $\langle
\sigma_{z} \rangle_{0} = 1$, then Eqs.~\eqref{eq:sigma-z}
and~\eqref{eq:sigma-z-square-average} tell us that both $\langle
\sigma_{z} \rangle_{t}$ and $\langle \sigma_{z} \rangle_{t}^2$
remain equal to 1 for each realization of the noise; in such a
case, the average value of $\exp ( i\gamma^{\makebox{\tiny
tot}}_{t}({\omega}))$ at time $T$ is:
\begin{equation}
{\mathbb E}_{\mathbb P} [ \exp ( i\gamma^{\makebox{\tiny
tot}}_{T}({\omega})) ] = e^{- \frac{\lambda^2}{2} \sin^2\! \varphi
\, T + i (\mu B + \lambda^2 \sin\varphi \cos\varphi) T},
\end{equation}
and its argument clearly depends on $\varphi$.

\subsection{Geometric phase}

The geometric phase $\gamma^{\makebox{\tiny geo}}_{t}$ is the
difference between the total and the dynamical phase. For each
realization of the stochastic process $W_{t}$, one has from
Eqs.~\eqref{eq:tot-ph} and~\eqref{eq:din-phase}\footnote{In
Eqs.~\eqref{eq:geo-phase} and~\eqref{eq:geo-phase-aver}, $t$
refers to the time during which the beam travels through the
interferometer, while $T$ is the time during which the spin
interacts with the magnetic field.}:
\begin{eqnarray} \label{eq:geo-phase}
\gamma^{\makebox{\tiny geo}}_{T}(\omega) & = & \makebox{Arg}[
f_{t}(\omega)] -  \mu B \int_{0}^{T} \langle \sigma_{z}
\rangle_{s}\, ds \\
& - & \!\lambda \sin\varphi\! \int_{0}^{T} \langle \sigma_{z}
\rangle_{s}\, dW_{s} \nonumber \\
& - & \lambda^2 \sin\varphi\cos\varphi \!\int_{0}^{T} \langle
\sigma_{z} \rangle_{s}^2 \, ds \nonumber
\end{eqnarray}
($f_{t}(\omega)$ is defined in~\eqref{eq:fom}), which clearly
depends on the type of unravelling. Its average value is:
\begin{equation} \label{eq:geo-phase-aver}
\gamma^{\makebox{\tiny geo}}_{T} = \makebox{Arg} {\mathbb
E}_{\mathbb P}[ f_{t}(\omega) ] - \left\{
\begin{array}{l}
{\mathbb E}_{\mathbb P}[ \gamma^{\makebox{\tiny dyn}}_{T}(\omega)]
\quad \makebox{or}\\
\\\makebox{Arg}[{\mathbb E}_{\mathbb P}[ \exp ( i\gamma^{\makebox{\tiny
dyn}}_{T}(\omega))]],
\end{array}
\right.
\end{equation}
Also the average geometric phase depends on the type of stochastic
unravelling of the master equation, whichever way the average
dynamical phase is computed. This is the main result of our paper.

\section{Discussion and conclusions}

The geometric phase of an open quantum system should be a quantity
depending only on the path followed by the density matrix
$\rho_{t}$ in its state space; we have seen that such a phase,
when computed by means of stochastic unravellings---as done in
ref. \cite{gen4}---depends on the type of unravelling, both for
single realizations of the noise (Eq.~\eqref{eq:geo-phase}) and
for its average value (Eq.~\eqref{eq:geo-phase-aver}). This fact
as two important consequences:
\begin{enumerate}
\item First of all, the phase defined in~\eqref{eq:geo-phase}
and~\eqref{eq:geo-phase-aver} is {\it not} a geometric object,
since it depends also on $\varphi$ which has nothing to do with
the path followed by $\rho_{t}$ during its evolution.

\item Worse than this, such a phase is not even an object somehow
related to a physical quantity, because $\varphi$ itself has no
physical meaning since it only selects one of the infinitely many
equivalent stochastic unravellings which can be used.
\end{enumerate}
The conclusion is that the stochastic-unravelling method does not
lead to a sensible definition of geometric phase.

One could say that different stochastic unravellings might
correspond to different ways to perform the measurement or to
monitor the environment \cite{Carmichael} and then that different
values for the phase correspond to different ways to measure the
system; anyway, this is not here the case: in our example we have
taken a standard interferometer where the output intensity is
measured in a standard and unique way. Nevertheless, different
unravellings can still be taken into account.

Note that the dependence on $\varphi$ comes only from the
dynamical phase, not from the total phase. As already remarked,
this is not a consequence of the specific model of open quantum
system we are considering here, but a direct consequence of the
fact that a total-phase difference is a measurable quantity and as
such must be deducible from the density matrix $\rho_{t}$, which
does not depend on $\varphi$. On the contrary, the dynamical phase
and thus also the geometrical phase, is not directly observable
so---at least from the mathematical point of view---it can depend
on $\varphi$, as it happens here.

What is the mathematical origin of the dependence of
$\gamma^{\makebox{\tiny geo}}_{t}(\omega)$ and
$\gamma^{\makebox{\tiny geo}}_{t}$ on $\varphi$? Its
unravelling-dependence does not come from the total phase but from
the dynamical component, and for the following reason: by
definition, $\gamma^{\makebox{\tiny dyn}}_{t}(\omega)$ is not a
function of $|\psi_{t}\rangle$ at the considered time, but a
function of the {\it whole} history of $|\psi_{s}\rangle$, from $s
= t_{0}$ to $s = t$, i.e. it depends on the whole trajectory
followed by the statevector. Now, it is easy to see that for
different unravellings the trajectories followed by the
statevector are radically different. For example, when $\varphi =
\pi/2 + k \pi$ with $k \in {\bf Z}$, Eq.~\eqref{eq:sigma-z} tells
that $\langle \sigma_{z} \rangle$ is constant in time, for {\it
each} realization of the stochastic process: this implies that the
projection of the spin vector along the magnetic field does not
change in time, i.e. the vector rotates always along the same
circle on the Bloch sphere. On the other hand, when $\varphi \neq
\pi/2 + k \pi$ with $k \in {\bf Z}$ then, as we have already
discussed in connection with
Eq.~\eqref{eq:sigma-z-square-average}, $\langle \sigma_{z}
\rangle^2$ approaches the value 1 for $t \rightarrow \infty$, i.e.
the variance $\makebox{Var} [\sigma_{z}] \equiv \langle
\sigma_{z}^2 \rangle - \langle \sigma_{z} \rangle^2 = 1 - \langle
\sigma_{z} \rangle^2$ of the operator $\sigma_{z}$ approaches
zero: this means that the statevector is driven towards one of the
two eigenstates of $\sigma_{z}$, thus changing the projection of
the spin vector along the magnetic field \cite{primo}. As a
consequence, being the trajectories followed by the statevector so
strongly dependent on the kind of unravelling, there is no need
for the dynamical phase to be unravelling independent, as it
actually occurs.

A different way to see what happens is the following: the relation
$\langle \psi_{t}| d | \psi_{t}\rangle = 0$ defines the parallel
transport condition, and for different unravellings one has
different inequivalent parallel transport conditions, thus
different definitions of a geometric phase.

Another interesting question is about the physical reason for such
a dependence of the geometric phase on $\varphi$. Someone
\cite{referee} has argued that the problem arises because the
master equation used to model the effect of the environment is of
the Lindblad type. Since the Lindblad equation is only an
effective equation approximating an otherwise too complex system,
its validity is limited and it could not be suitable for computing
the geometric phase. We think that this is not the case: the
source of all troubles derives from the fact that there are
different equivalent stochastic unravellings associated to the
same evolution $\Sigma_{t}$ which determines different evolutions
for the statevector, thus different parallel transport conditions;
such a feature is not an exclusive property of the Lindblad
equation (it is not even a mathematical consequence of it), but
has a more general character.

To summarize, the stochastic-unravelling approach used in
\cite{gen4} does not produce a phase which is geometric, i.e.
which depends only on the trajectory followed by $\rho_{t}$ during
the evolution; it depends also on the specific choice of the
unravelling used for the calculations, which by itself has no
particular physical meaning. This difficulty can be in principle
overcome by fixing the unravelling to be used for computing the
geometric phase, as often implicitly done in the literature, but
this procedure cannot be satisfactory for two reasons: first, it
obviously does not remove the fact that the definition is
mathematically unravelling-dependent; second, there is no
fundamental physical reason why to choose one unravelling in place
of another, since they are all on the same footing.

In ref. \cite{new} it has been stated that, within the state
purification approach of \cite{gen3.1}, different Kraus
representations may lead to different values for the geometric
phase; if so, then our criticism apply also to the approach of
\cite{gen3.1}.

\section{Acknowledgements}

The work of A.B. was supported by the Marie Curie Fellowship
MEIF-CT-2003-500543. The work of E. I. was supported by the
``Consorzio per la Fisica --- Trieste''.

\end{document}